\documentclass[useAMS,usenatbib,babel]{aa}

\def\gtrsim{\lower.5ex\hbox{$\; \buildrel > \frac \sim \;$}}

\usepackage{color}

\begin{document}

\title{An implicit scheme for solving the anisotropic diffusion of heat and cosmic rays in the RAMSES code}

\titlerunning{Implicit scheme for anisotropic diffusion}
\authorrunning{Y. Dubois \& B. Commer\c con}

\author{Yohan Dubois\inst{1} \and Beno\^it Commer\c con\inst{2}}
\institute{Institut d'Astrophysique de Paris, UMR 7095, CNRS, UPMC Univ. Paris VI, 98 bis boulevard Arago, 75014 Paris, France \\\email{dubois@iap.fr} \and \' Ecole Normale Sup\'erieure de Lyon, CRAL, UMR 5574 du CNRS, Universit\'e de Lyon I, 46 All\'ee d'Italie, 69364 Lyon Cedex 07, France \\ \email{benoit.commercon@ens-lyon.fr}} 

\date{Received / Accepted }

\abstract{Astrophysical plasmas are subject to a tight connection between magnetic fields and the diffusion of particles, which leads to an anisotropic transport of energy. Under the fluid assumption, this effect can be reduced to an advection-diffusion equation, thereby augmenting the equations of magnetohydrodynamics. We introduce a new method for solving the anisotropic diffusion equation using an implicit finite-volume method with adaptive mesh refinement and adaptive time-stepping in the {\sc ramses} code. We apply this numerical solver to the diffusion of cosmic ray energy and diffusion of heat carried by electrons, which couple to the ion temperature. We test this new implementation against several numerical experiments and apply it to a simple supernova explosion with a uniform magnetic field.}


\maketitle

\section{Introduction}

Diffusive processes are relevant at various scales in astrophysical systems from the diffusion of cosmic rays (CRs) in the interstellar medium~\citep[e.g.][]{strongetal07} to the the diffusion of heat and CRs in cluster of galaxies~\citep[e.g.][]{rosner&tucker89, fabian94, narayan&medvedev01}.
As the diffusion of these charged particles operate in a magnetised plasma, the diffusion process becomes anisotropic since it is conducted along magnetic field lines.
Furthermore, for a fully ionised plasma, like the lighter population of particles, electrons conduct heat much faster than protons, which are virtually in the limit of non-diffusion.
In some situations, it leads to a non-thermal equilibrium between ions and electrons if the collisional coupling time of the two temperatures is longer than the diffusion timescale of electrons.
Thus, there could be fluctuations in electron temperature without the corresponding fluctuation in the ion temperature.

Those processes have mostly been ignored from the numerical astrophysics perspective with a few exceptions.
In the intra-cluster medium, it has been shown that ions and electrons are not perfectly at thermal equilibrium~\citep{chiezeetal98, courty&alimi04, rudd&nagai09, wong&sarazin09, gaspari&churazov13} and that the transport of heat produces an electron precursor at the virial radius of massive dark matter halos~\citep{teyssieretal98}. The transport of heat eventually reduces the cooling flow at the core of galaxy clusters, transporting the cosmic shock-driven internal energy of the outskirts towards the centre~\citep{ruszkowski&begelman02, jubelgasetal04}.
Thermal and magnetohydrodynamical instabilities form because of the anisotropic nature of the conduction, the so-called magneto-thermal instability~\citep{balbus00, parrishetal08} and the heating-buoyancy instability~\citep{parrish&quataert08, bogdanovicetal09, parrishetal09}, and their variations with the adjunction of the diffusion of CRs~\citep{rasera&chandran08}.
Conduction may also play an important role in redistributing the mechanical energy deposited by active galactic nuclei jets~\citep{ruszkowski&begelman02, brighenti&mathews03, bruggenetal05, mcnamara&nulsen07}, helping to solve the cooling catastrophe in galaxy clusters.
A similar heating source in the intra-cluster medium is that of comic rays deposited in active galactic nuclei or at shocks and redistributed by diffusion~\citep{miniatietal01, ensslinetal11}.

On galactic scales, since CR energy in the interstellar medium (ISM) is at equipartition with kinetic, thermal, and magnetic energies~\citep{beck&krause05}, they may play a crucial role in the self-regulation of the ISM dynamics and, thus, of star formation.
For instance, the injection of CRs within remnants of supernovae lead to stronger galactic-scale winds~\citep{jubelgasetal08, uhligetal12, hanaszetal13, boothetal13, salem&bryan14} and to amplified galactic dynamos with anisotropic diffusion~\citep{hanaszetal04}.
Anisotropic conduction in the ISM affects the shape and size of cold clouds~\citep{koyama&inutsuka04, piontek&ostriker04, choi&stone12}, and the expansion of supernova remnants~\citep{tilleyetal06, balsaraetal08sn}.
Also, CRs can penetrate deep inside cold dense cores and, as a consequence, can regulate the ionisation rates of the gas~\citep[e.g.][]{spitzer&tomasko68, padovanietal09}.

One problem with the implementation of diffusion in numerical simulation is that the stability criterion is $\Delta t_{\rm diff}= \Delta x^2/(2D_{\rm diff})$, where $\Delta x$ is the cell size and $D_{\rm diff}$ the diffusion coefficient.
For comparison, the hydrodynamical Courant Friedrich Levy (CFL) condition is $\Delta t_{\rm h}=C \Delta x / (u+c_{\rm s})$, where $u$ is the gas velocity, $c_{\rm s}$  the fluid sound speed, and $C\le 0.8$ is the Courant factor.
Since the diffusion stability time step does not scale linearly with resolution, unlike the hydrodynamical time step, it could be a bottleneck to employ an explicit diffusion scheme that must verify this time step condition at any time, especially for multi-scale problems where gravity shapes strong contrasts in gas densities and triggers refinements in resolution.

Implicit numerical solvers in these situations could be favoured.
They do not need to fulfil the time step constraint on diffusion, and any time step can be chosen (such as the CFL time step), at the expense of numerical complexity and intensity compared to an explicit solver.
There is a variety of numerical implementations of anisotropic diffusion from centred symmetric to centred asymmetric schemes~\citep{gunteretal05} using various slope limiters to preserve the monotonicity of the system~\citep{sharma&hammett07}.
Amongst all these implementations, implicit and explicit solvers have been developed for astrophysical codes: \cite{hanaszetal03}, \cite{parrish&stone05}, and \cite{rasera&chandran08} implemented an explicit method for anisotropic conduction and diffusion, \cite{balsaraetal08method} developed a semi-implicit solver, \cite{yokoyama&shibata01} and \cite{meyeretal12} a fully implicit solver.

In this paper, which uses the {\sc ramses} code~\citep{teyssier02}, we present the first implementation of an implicit solver for anisotropic diffusion or conduction on adaptive mesh refinement (AMR) with adaptive time-stepping.
This new numerical implementation is augmented by modelling a multi-temperature and energy component with temperature coupling of ions and electrons.
In section~\ref{section:numerics}, we present the new numerical solver for diffusion or conduction and temperature coupling.
In section~\ref{section:tests}, we test our method with several numerical experiments.
We discuss future developments and applications of this method in section~\ref{section:conclusion}.

\section{Numerical set-up}
\label{section:numerics}

\subsection{Magnetohydrodynamics with heat and cosmic ray diffusion}

The set of differential equations to be solved for magnetohydrodynamics of a fluid made of ions and electrons with heat conduction and CRs diffusion is 
\begin{eqnarray}
\label{mass} \frac{\partial \rho}{\partial t} &+& \nabla. (\rho \vec{u})= 0 \, ,\\
\label{momentum}\frac{\partial \rho \vec{u}} {\partial t} &+& \nabla. \left(\rho \vec{u}\vec{u}+p_{\rm tot}-\frac{\vec{BB}}{4 \pi}\right)= 0\, ,\\
\label{energy}\frac{\partial e} {\partial t} &+& \nabla. \left((e+p_{\rm tot})\vec{u}-\frac{\vec{B (B.u)}}{4 \pi}\right) \nonumber \\ 
&=&-\nabla.\vec{F_{\rm cond}}- \nabla.\vec{F_{\rm CR}} \, ,\\
\label{magnetic}\frac{\partial \vec{B}} {\partial t} &-& \nabla \times (\vec{u} \times \vec{B} )=0\, ,\\
\label{electron}\frac{\partial e_{\rm E}} {\partial t} &+& \nabla. (e_{\rm E}\vec{u})=-p_{\rm E}\nabla.\vec{u} - \nabla.\vec{F_{\rm cond}} + \mathcal H_{\rm EI} \\
\label{cosmic}\frac{\partial e_{\rm cr}} {\partial t} &+& \nabla. (e_{\rm cr}\vec{u})=-p_{\rm cr}\nabla.\vec{u} - \nabla.\vec{F_{\rm CR}}\, ,
\end{eqnarray}
where $\rho$ is the gas mass density, $\vec{u}$  the gas velocity, $\vec B$  the magnetic field, $e=0.5\rho u^2+e_{\rm th}+e_{\rm cr}+ B^2/8\pi$ is the total energy density, $e_{\rm th}=e_{\rm I}+e_{\rm E}$ is the thermal energy density, $e_{\rm E}$ and $e_{\rm I}$ are the electron and ion energy densities, $e_{\rm cr}$ is the energy density of CRs, and total pressure $p_{\rm tot}=(\gamma-1)e_{\rm th}+(\gamma_{\rm cr}-1)e_{\rm cr}+0.5B^2/4\pi$, where $\gamma$ and $\gamma_{\rm cr}$ are the adiabatic indexes of the gas and CRs, respectively.
All energy components $e_{i}$ are energies per unit volume $e_{i}=E_{i}/\Delta x^3$, where $\Delta x$ is the cell size.
The heat flux $\vec{F_{\rm cond}}$ is carried by electrons alone, and the temperature of ions and electrons couple through the energy exchange term $\mathcal H_{\rm EI}$. 
Since ions are heavier than electrons, their thermal velocity is lower by a factor $\propto \sqrt{m_{\rm E}/m_{\rm I}}$.
Therefore, for a given increase in thermal velocity at a shock, it corresponds to a higher temperature increase for ions than for electrons.
It justifies the fact that electron energy is treated in a separate energy equation that does not fulfil the jump condition.
The same arguments apply to CRs compared to the thermal gas since velocities of CRs are much higher than thermal velocities of the ions (a.k.a. CRs are a non-thermal gas component).

We assume that the thermal components of the gas are made only of ions and electrons, meaning that the gas is fully ionised.
In some situations, this approximation will break down (star-forming regions), but it is a good approximation for the warm or hot phases of the ISM and for the intergalactic gas.
CRs also diffuse through the diffusion flux term $\vec{F_{\rm CR}}$.
Equation~(\ref{cosmic}) can be repeated for as many CR energy components required to decompose the CR energy power spectrum (with different diffusion coefficients). 
Introducing multiple CR energy bands requires including adiabatic and radiative losses for a proper treatment of the CR power spectrum, so we defer this to future work and simply assume that CRs are represented by one fluid with one effective diffusion coefficient and one effective adiabatic index.

We use the AMR {\sc ramses} code detailed in~\cite{teyssier02}.
Equations~(\ref{electron}) and~(\ref{cosmic}) for electrons and CRs, respectively, are added to the set of MHD equations.
The full set of equations is solved with the standard MHD solver of {\sc ramses} described in~\cite{fromangetal06}, where the right-hand terms of equation~(\ref{energy}) are treated separately as source terms. 
The ion internal energy is obtained for free by subtracting the electron internal energy from the total thermal energy.
The induction equation~(equation \ref{magnetic}) is solved using constrained transport~\citep{teyssieretal06}.
Godunov fluxes are modified to account for the extra energy components and total pressure made of ions, electrons, and CRs.
Therefore, the effective sound speed used for the CFL time step condition accounts for the extra pressure components (i.e. total pressure of the fluid).

\subsection{Anisotropic diffusion}

We explain the method for the conduction of heat through electrons, but this can be applied to a single temperature model (ions and electrons at thermal equilibrium) or for the diffusion of CRs.
The difference for CRs is that the diffusion coefficient $D_{\rm CR}$ is constant throughout the simulation volume and that it applies to the gradient of CR energy instead of the gradient of temperature, e.g. $\vec F_{\rm CR}=-D_{\rm CR} \vec b (\vec b .  \nabla) e_{\rm CR}$, where $\vec b$ is the magnetic field unit vector.

In the presence of magnetic fields, the conduction of heat in a plasma is 
\begin{equation}
\label{eq:conduction}\frac{\partial e_{\rm E}} {\partial t}=-\nabla .\vec F_{\rm cond}=-\nabla.\left(-\kappa_{\parallel}\vec b (\vec b.\nabla) T_{\rm E} \right)-\nabla.\left(-\kappa_{\rm iso}\nabla T_{\rm E} \right)\, ,
\end{equation}
where $\kappa_{\rm iso}$ and $\kappa_{\parallel}$ are the isotropic conduction and parallel along the magnetic field lines coefficients respectively, with $\kappa_{\parallel}=\kappa_{\rm Sp}-\kappa_{\rm iso}$, and $\kappa_{\rm iso}=f_{\rm iso}\kappa_{\rm Sp}$ with $f_{\rm iso}$ an arbitrary small coefficient for anisotropic diffusion ($f_{\rm iso}=1$ is for isotropic diffusion).
In many astrophysical cases, $\kappa_{\rm iso}/\kappa_{\parallel}\ll1$ since the Larmor radius is much smaller than the mean free path of electrons.
For instance, in the hot gas of galaxy clusters with temperature $T_{\rm E}=3$ keV,
electron density $n_{\rm E}=10^{-2}\, \rm cm^{-3}$ and $B=1\, \rm \mu G$, the Larmor radius is $\lambda_{\rm L}=10^8\, \rm cm$ and the mean free path $\lambda_{\rm mfp}=10^{21}\, \rm cm$. 
However, to ensure numerical stability, we set up the isotropic conduction term to be around one per cent of the parallel conduction coefficient.

The conduction coefficient for electrons is the~\cite{spitzer56} value 
\begin{equation}
\kappa_{\rm Sp}=n_{\rm E}k_{\rm B} D_{\rm C}\, ,
\end{equation}
where $k_{\rm B}$ is the Boltzmann constant and $D_{\rm C}$ the thermal diffusivity
\begin{equation}
D_{\rm C}=8\times 10^{31} \left( \frac{T_{\rm E}}{10 {\rm keV}}\right )^{\frac{5}{2}}\left( \frac{n_{\rm E}}{5\times 10^{-3} \, \rm cm^{-3}}\right)^{-1} \, \rm cm^2\, s^{-1}\, .
\end{equation}

\subsubsection{Implicit scheme}

This diffusion step is performed after the magnetohydrodynamics step. 
Here, we explain how the anisotropic conduction and diffusion equation is solved for a uniform grid, and the AMR part of the solver is explained in section~\ref{section:amr}.
Given the restrictive stability condition for an explicit scheme on the equation~(\ref{eq:conduction}), $\Delta t_{\rm diff}= \Delta x^2/(2D)$, compared to the behaviour of the CFL condition for the hydrodynamics time step $\Delta t_{\rm h}=\Delta x / (u+c_{\rm s})$, using an implicit solver on the equation diffusion will alleviate this numerical stability constraint.
Thus, the diffusion term is solved over one hydrodynamical time step using implicit integration.
Discretising equation~(\ref{eq:conduction}) in 1D leads to (we omit subscripts E and cond for clarity)
\begin{equation}
e^{n+1}_{i}+\Delta t \frac{F^{n+1}_{i+\frac{1}{2}} - F^{n+1}_{i-\frac{1}{2}}} {\Delta x}= e^{n}_{i}\, ,
\end{equation}
where the subscript $i$ is for the cell position, and the superscript $n$ for the time.
Each of the fluxes $F^{n+1}$ is evaluated at the cell interface (2 in 1D, 4 in 2D, and 6 in 3D) at time $n+1$.
This can be rewritten as
\begin{equation}
\label{eq:discr}
e^{n+1}_{i}-\Delta t \frac{\kappa^{n}_{i+\frac{1}{2}}(T^{n+1}_{i+1} -T^{n+1}_{i}) - \kappa^{n}_{i-\frac{1}{2}}(T^{n+1}_{i} -T^{n+1}_{i-1})} {\Delta x^2}= e^{n}_{i}\, ,
\end{equation}
where $e_i=n_i k_{\rm B} T/(\gamma-1)$ ($n_i$ here is the gas number density, not time index $n$). Equation \ref{eq:discr} implicitly assumes that the cells are cubic, which is the case in {\sc ramses}.
We use the value of $\kappa$ at time $n$ since we employ an implicit solver, which requires the constancy of the coefficient during the integration step.
Thus, defining $C_{i\pm\frac{1}{2}}=\kappa_{i\pm\frac{1}{2}}\Delta t/\Delta x^{2} $ and $C_{{\rm v},i}= n_i k_{\rm B}/(\gamma-1)$, we obtain
\begin{equation}
\label{eq:matrix}-C^{n}_{i-\frac{1}{2}}T^{n+1}_{i-1} + (C_{{\rm v}, i}^n+C^{n}_{i-\frac{1}{2}}+C^{n}_{i+\frac{1}{2}}) T^{n+1}_{i} -C^{n}_{i+\frac{1}{2}}T^{n+1}_{i+1}= C_{{\rm v}, i}^nT^{n}_{i}\, .
\end{equation}

We discretise equation~(\ref{eq:conduction}) in 2D (3D can be obtained from 2D with little effort) assuming the cells have the same extent in $x$, $y$, $z$ directions
\begin{equation}
\label{eq:2dcond}e^{n+1}_{i,j}+\Delta t \frac{F^{n+1}_{i+\frac{1}{2},j}+F^{n+1}_{i,j+\frac{1}{2}} - F^{n+1}_{i-\frac{1}{2},j} - F^{n+1}_{i,j-\frac{1}{2}}} {\Delta x}= e^{n}_{i,j}\, ,
\end{equation}
for cell position $i,j$. These quantities are evaluated with the centred symmetric scheme proposed by~\cite{gunteretal05} for the anisotropic part of the flux.
The anisotropic flux at cell interfaces $F^{\rm ani}_{i\pm1/2,j}$ and $F^{\rm ani}_{i,j\pm1/2}$ are evaluated from their cell corner fluxes $F^{\rm ani}_{i\pm1/2,j\pm1/2}$, thus
\begin{eqnarray}
F^{\rm ani}_{i+\frac{1}{2},j}&=&\frac {F^{\rm ani}_{i+\frac{1}{2},j-\frac{1}{2}} + F^{\rm ani}_{i+\frac{1}{2},j+\frac{1}{2}}} {2}\, , \nonumber \\
F^{\rm ani}_{i,j+\frac{1}{2}}&=&\frac {F^{\rm ani}_{i-\frac{1}{2},j+\frac{1}{2}} + F^{\rm ani}_{i+\frac{1}{2},j+\frac{1}{2}}} {2}\, . \nonumber 
\end{eqnarray}
The anisotropic cell corner flux is 
\begin{equation}
\label{eq:aniflux}F^{\rm ani}_{i+\frac{1}{2},j+\frac{1}{2}}=\bar \kappa_{\parallel} \bar b_{x} \left( \bar b_{x} \bar {\frac{\partial T} {\partial x}} + \bar b_{y} \bar{ \frac{ \partial T} {\partial y} }\right)\, ,
\end{equation}
where barred quantities are arithmetic averages over the cells connected to the corner; i.e.,
\begin{eqnarray}
\bar b_{x} &=& \frac {b^n_{x,i+\frac{1}{2},j}+b^n_{x,i+\frac{1}{2},j+1}} {2} \, , \nonumber \\
\bar b_{y} &=& \frac {b^n_{y,i,j+\frac{1}{2}}+b^n_{y,i+1,j+\frac{1}{2}}} {2} \, , \nonumber \\
\bar {\frac {\partial T} {\partial x}} &=& \frac {T^{n+1}_{i+1,j+1} + T^{n+1}_{i+1,j} -T^{n+1}_{i,j+1} -T^{n+1}_{i,j}} {2\Delta x} \, , \nonumber \\
\bar {\frac {\partial T} {\partial y}} &=& \frac {T^{n+1}_{i+1,j+1} + T^{n+1}_{i,j+1} -T^{n+1}_{i+1,j} -T^{n+1}_{i,j}} {2\Delta x} \, , \nonumber \\
\bar \kappa_{\parallel} &=& \frac{\kappa^n_{\parallel i,j}+\kappa^n_{\parallel i+1,j}+\kappa^n_{\parallel i,j+1}+\kappa^n_{\parallel i+1,j+1}} {4} \, .\nonumber 
\end{eqnarray}

For the isotropic part of the flux, we use a classical discretisation, where the fluxes of equation~(\ref{eq:2dcond}) are simply obtained from the left and right states of the cell interface; i.e., 
\begin{equation}
F^{\rm iso}_{i+\frac{1}{2}, j}=\frac{\kappa^n_{{\rm iso}\,  i+1, j}+\kappa^n_{{\rm iso} \, i, j}} {2} \frac{T^{n+1}_{i+1, j}-T^{n+1}_{i, j}} {\Delta x}\, .
\end{equation}

With this numerical scheme, the matrix system $A\vec{x}=\vec{c}$ formed by equation~(\ref{eq:matrix}), with $A$ the matrix including the $C^n$ and $C_{{\rm v}}^n$ coefficients, $\vec{x}$ the vector of temperature $T^{n+1}$, and $\vec{c}$ the vector of energy densities $C_{{\rm v}}^nT^n$ can be generalised in multi-dimensions by equation~(\ref{eq:2dcond}).
The sparse matrix $A$ is positive definite and symmetric, so we can use the conjugate gradient algorithm to solve this system of linearised equations as in~\cite{commerconetal11} (for radiation hydrodynamics in their case).

\subsubsection{Diffusion with adaptive mesh refinement}
\label{section:amr}

We follow the procedure introduced in~\cite{commerconetal14} for solving the diffusion equation with AMR and adaptive time-stepping on a level-by-level basis.
Each level $\ell$ is evolved with a time step $\Delta t^\ell$ that is twice less than the coarser level $\ell-1$, $\Delta t^\ell =\Delta t^{\ell -1}/2$, meaning that level $\ell$ evolves with two consecutive time steps before doing one time step of level $\ell-1$. 
The adaptive time-stepping is performed for all solvers in {\sc ramses} including the diffusion solver presented in this paper, and finer levels are updated first.

At a given level of refinement $\ell$, there are two types of non-uniform interfaces: the fine-to-coarse interface (interface between a cell at level $\ell$ and a cell at $\ell -1$) and the coarse-to-fine interface (interface between a cell at level $\ell$ and a cell at $\ell +1$).
In both cases, we use Dirichlet boundary conditions: cell values at level boundaries are imposed. One could choose Neumann boundary conditions, which impose the fluxes and guarantee energy conservation as what is done for the hydrodynamical solver. However, as shown in~\cite{commerconetal14}, this sort of imposed flux conditions could lead to negative values for lang time steps.  

For the fine-to-coarse interface, we use values of $\ell-1$ at time $n$ as imposed boundary conditions for level $\ell$.
With the minmod scheme~\citep{vanleer79} (monotonised central is also a valid choice), we interpolate the values of level $\ell -1$ on a finer virtual grid at level $\ell$ to impose the fine-to-coarse boundary.
For the coarse-to-fine interface, we use values of $\ell+1$ at time $n+1$ as imposed boundary conditions for level $\ell$.
We restrict the value of the boundary coarse cell at level $\ell$
 to the average value of the $2^{dim}$ cells of level $\ell+1$ 
 to impose the coarse-to-fine boundary.
In any case, eq. \ref{eq:2dcond} remains correct since the level-by-level diffusion solver only deals with data estimated at the same level of refinement. The combination of the use of Dirichlet boundary conditions at level interfaces and of interpolation or restriction operations does not break the symmetry of matrix $A$.  The imposed values of the neighbouring cells at different refinement levels go into the right-hand side (vector $\vec c$) of the matrix system $A\vec x=\vec c$, because it is the case for the computational domain imposed boundary conditions.

\subsubsection{Limitations of the method}

There are two main limitations to our current anisotropic diffusion solver: the non-energy conservation due to Dirichlet boundary at level interfaces and the non-monotonicity preserving nature of the solver.
To circumvent the first limitation, one could adopt a unique time step strategy, where diffusion is solved for all levels at once.
This guarantees energy conservation but slows down the calculation.
A good compromise would be to do the diffusion step rather than every fine time step of the simulation, but only over coarse time steps (or even every 10, 100, etc. coarse time steps).
Nevertheless, as shown in~\cite{commerconetal14}, Dirichlet boundary conditions are robust in most cases, and in practice, energy conservation is acceptable (see tests below).

Concerning the non-monotonicity preserving nature of the diffusion solver, since the flux at a cell interface (equation~\ref{eq:aniflux}) includes a transverse gradient of temperature, it is not guaranteed that the flux flows from the high temperature cell to the low temperature cell.
Thus, the temperature can become negative, and the monotonicity of the solution is not preserved (see for instance the test problem illustrated in figure 5 of \citealp{sharma&hammett07}).
\cite{sharma&hammett07} propose a slope limiter for the centred symmetric scheme with explicit time integration that preserves the monotonicity of the solution.
However, employing a slope limiter would not allow us to use the conjugate gradient algorithm any longer since matrix $A$ becomes non-symmetric.
In our case, we prefer to conserve the simple and fast solution offered by the conjugate gradient algorithm and to fix the negative temperatures created by the anisotropic conduction solver with tiny positive values.
Future developments will include slope limiters, and the matrix system will be solved using a bi-conjugate gradient algorithm instead (as in~\citealp{gonzalezetal15}).

\subsection{Ion and electron temperature coupling}
\label{section:coupling}

We now explain how we model the coupling of the temperature of ions and electrons in {\sc ramses}.
This coupling step is performed after the diffusion step.
The energy exchange rate $\mathcal H_{\rm EI}$ introduced in equation~(\ref{electron}) is written for a gas that we assume to be fully ionised and composed of hydrogen and helium
\begin{eqnarray}
\frac{de_{\rm E}} {dt}&=&\mathcal H_{\rm EI}= \frac{ T_{\rm I}-T_{\rm E}} {\tau_{\rm eq, EI}} \frac{n_{\rm E}k_{\rm B}} {\gamma-1} \, ,\nonumber \\
\frac{de_{\rm I}} {dt}&=&- \mathcal H_{\rm EI}=\frac{ T_{\rm E}-T_{\rm I}} {\tau_{\rm eq, IE}} \frac{n_{\rm I}k_{\rm B}} {\gamma-1}\, , \label{eqion}
\end{eqnarray}
with the equilibration timescales
\begin{eqnarray}
\tau_{\rm eq, EI}&=&\frac{3 m_{\rm E}m_{\rm I}k_{\rm B}^{\frac{3}{2}}} {8 (2 \pi)^{\frac{1}{2}} n_{\rm I} Z_{\rm I}^2 q_{\rm E}^4 \ln \Lambda} \left( \frac{T_{\rm E}} {m_{\rm E}} + \frac{T_{\rm I}} {m_{\rm I}} \right)^{\frac{3}{2}} \nonumber \\
&\simeq&\frac{3 m_{\rm E}m_{\rm I}k_{\rm B}^{\frac{3}{2}}} {8 (2 \pi)^{\frac{1}{2}} n_{\rm I} Z_{\rm I}^2 q_{\rm E}^4 \ln \Lambda} \left( \frac{T_{\rm E}} {m_{\rm E}} \right)^{\frac{3}{2}}\, ,
\end{eqnarray}
and
\begin{eqnarray}
\tau_{\rm eq, IE}&=&\frac{3 m_{\rm E}m_{\rm I}k_{\rm B}^{\frac{3}{2}}} {8 (2 \pi)^{\frac{1}{2}} n_{\rm E} Z_{\rm I}^2 q_{\rm E}^4 \ln \Lambda} \left( \frac{T_{\rm E}} {m_{\rm E}} + \frac{T_{\rm I}} {m_{\rm I}} \right)^{\frac{3}{2}} \nonumber \\
&\simeq&\frac{3 m_{\rm E}m_{\rm I}k_{\rm B}^{\frac{3}{2}}} {8 (2 \pi)^{\frac{1}{2}} n_{\rm E} Z_{\rm I}^2 q_{\rm E}^4 \ln \Lambda} \left( \frac{T_{\rm E}} {m_{\rm E}} \right)^{\frac{3}{2}}\, ,
\end{eqnarray}
where $m_{\rm E}$ and $m_{\rm I}$ stand for the electron and ion mass, respectively; $n_{\rm I}=\rho /(\mu_{\rm I}m_{\rm p})$ is the ion number density with $m_{\rm p}$ the proton mass and $\mu_{\rm I}$ being the ion mean molecular weight; $Z_{\rm I}$ is the ion charge number; $q_{\rm E}$  the electron charge; and $\ln \Lambda=40$ is the Coulomb logarithm.
Since $m_{\rm E}\ll m_{\rm I}$, we neglect the $T_{\rm I}/m_{\rm I}$ term in the energy exchange rates.
For a mixture of hydrogen and helium, the timescale of the temperature coupling between hydrogen and helium would be proportional to $m_{\rm H, He}^{-3/2}\ll m_{\rm E}^{-3/2}$.
Thus, the equilibration between hydrogen and helium operates $10^{5}$ times faster than the equilibration timescale between hydrogen and electrons or helium and electrons.
For hydrogen or helium, the ratio of $m_{\rm H, He}/Z^2_{\rm H, He}$ also simplifies to $m_{\rm p}$ (not the case for heavier elements).
Therefore, protons and HeIII can be considered as one single population sharing the same temperature.

Writing $\tilde \tau_{\rm eq}=\tau_{\rm eq, EI} / n_{\rm E}=\tau_{\rm eq, IE} / n_{\rm I}$, we see that 
\begin{equation}
\frac{de_{\rm E}} {dt}= -\frac{de_{\rm I}} {dt}= \frac{T_{\rm I}-T_{\rm E}} {\tilde \tau_{\rm eq}} \frac{k_{\rm B}}{\gamma-1}\, .
\end{equation}
The variation in energy is symmetric between the transfer of ion energy and electron energy, but this is not the case for the variation in temperature, which is $\partial_{t} T_{\rm E,I}\propto  n_{\rm E,I}^{-1}\, \partial_{t}  e_{\rm E,I}$. Therefore, the difference in temperature change between ions and electrons will be a factor $\mu_{\rm I}/\mu_{\rm E}$, where $\mu_{\rm E}$ is the ``mean molecular weight of electrons'' $\mu_{\rm E}=\rho/(n_{\rm E}m_{\rm p})$. For astrophysical plasmas essentially composed of hydrogen and helium, the change in temperature between ions and electrons is close to symmetric.

To solve for the system of two non-linear coupled equations, we rewrite~(\ref{eqion}) as 
\begin{eqnarray}
G_1(T^{n+1}_{\rm E}, T^{n+1}_{\rm I})&=&\Delta t \left [ -(T^{n+1}_{\rm E})^{-\frac{1}{2}}+T^{n+1}_{\rm I} (T^{n+1}_{\rm E})^{-{\frac{3}{2}}}\right ] \nonumber \\
&-& K \left [ T^{n+1}_{\rm E}-T^{n}_{\rm E} \right ] = 0 \, , \nonumber\\
G_2(T^{n+1}_{\rm E}, T^{n+1}_{\rm I})&=&\Delta t \left [ (T^{n+1}_{\rm E})^{-\frac{1}{2}}-T^{n+1}_{\rm I} (T^{n+1}_{\rm E})^{-{\frac{3}{2}}}\right ] \nonumber \\
&-& K \left [ T^{n+1}_{\rm I}-T^{n}_{\rm I} \right ] = 0 \, ,
\label{gzero}
\end{eqnarray}
where $K={3 m_{\rm p}k_{\rm B}^{1.5} / [8 (2 \pi m_{\rm E})^{{0.5}} n_{\rm E}  q_{\rm E}^4 \ln \Lambda}]$.
We introduce the vector $X=(T^{n+1}_{\rm E},T^{n+1}_{\rm I})^T$, and rewrite the previous two equations as
\begin{eqnarray}
G_1(X^k+\Delta X)-G_1(X^k)&=&\left (\frac{\partial G_1} {\partial X}\right)^k \Delta X\, , \nonumber \\
G_2(X^k+\Delta X)-G_2(X^k)&=&\left (\frac{\partial G_2} {\partial X}\right)^k \Delta X\, .\nonumber
\end{eqnarray}
The goal is now to iterate on $\Delta X$ to find the value of $G_1(X^k+\Delta X)=0$ and $G_2(X^k+\Delta X)=0$ as required by equations~(\ref{gzero}).
The derivatives $\partial_X G_1^k$ and $\partial_X G_2^k$ are obtained by differentiating equations~(\ref{gzero}) with respect to $T^{k}_{\rm E}$ and $T^{k}_{\rm I}$, where the superscript $k$ replaces the superscript $n+1$ in~(\ref{gzero}).
The values of $\Delta X$ are obtained with Cramer's rule, and $\Delta X$ is added to the value of $X^k$ until the variation in $\Delta X/ X^k<10^{-4}$.

\section{Numerical tests}
\label{section:tests}

If not indicated, we use a courant factor of $C=0.8$ with adaptive time-stepping between the different levels of refinement.

\subsection{Diffusion of a step function}

\begin{figure}
\centering \includegraphics[width=0.4\textwidth]{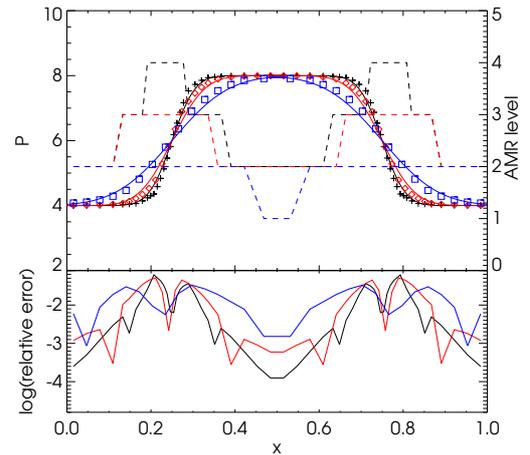}
\caption{Top: Diffusion of a pressure step function as a function of position at $t=9.3\times 10^{-4}$ (black), $t=1.9\times 10^{-3}$ (red), $t=5.6\times10^{-3}$ (blue) for the simulation (symbols) and the analytic solution (solid lines). The extra levels of refinement are indicated as dashed lines. Bottom: Relative errors of the simulation with respect to the analytic solution. }
\label{fig:step}
\end{figure}

We performed a simple 1D test of the diffusion of a step function with a constant diffusion coefficient of $D=1$ without hydrodynamics.
The initial setup is a pressure of $P_1=0.8$ between $x=[0.25,0.75]$ and $P_0=0.4$ outside with $C_{\rm V}=\rho k_{\rm B}/(\mu m_{\rm p})=1$.
We start with a minimum level of refinement of $\ell_{\rm min}=3$ and refine up to level $\ell_{\rm max}=7$ wherever the relative pressure variation is larger than $10\, \%$.
The time step is chosen to be $\Delta t_{7}=7\times \Delta t_{\rm diff}$ (for level 7), where $\Delta t_{\rm diff}$ is the stability time step for diffusion (only useful for explicit schemes). 

The time evolution of the step function has the following analytical solution for a constant diffusion coefficient
\begin{equation}
P(x,t)=P_0+\frac{P_1-P_0}{2}{\rm erf} \left( \frac{x-x_0}{\sqrt{4Dt}} \right) \, ,
\end{equation}
where $x_0=0.25$.
Figure~\ref{fig:step} shows the result of the diffusion of the step function with the analytical solution at three different times.
The analytical solution is reproduced well, below a $10\%$ relative error.

\subsection{Temperature coupling}

\begin{figure}
\centering \includegraphics[width=0.4\textwidth]{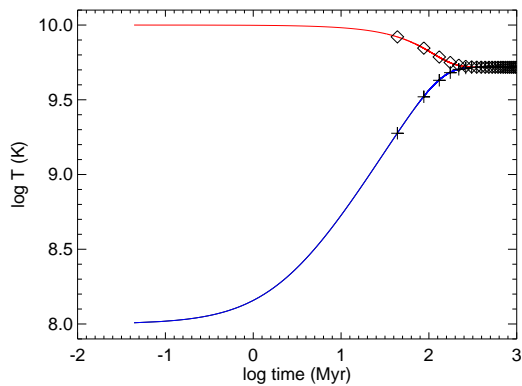}
\caption{Electron temperature (red or diamonds) and ion temperature (blue or pluses) as a function of time for the one cell problem. The solution for a time step of $\tau_{ \rm eq, EI,0}/50$ is represented in solid lines and for the time step of $20\tau_{ \rm eq, EI,0}$ with symbols.}
\label{fig:coupling}
\end{figure}

Figure~\ref{fig:coupling} shows the result of the one cell problem with two different initial temperatures for electrons $T_{\rm E,0}=10^{10}\, \rm K$ and ions $T_{\rm I,0}=10^{8}\, \rm K$.
The solid lines indicate the solution with a time step of $\tau_{ \rm eq, EI,0}/50$, and points are the solution with a time step of $20\tau_{ \rm eq, EI,0}$.
They are in excellent agreement, even for the run where the time step is 20 times the timescale of temperature coupling.

\subsection{Sod test with two energy components}

\begin{figure}
\centering \includegraphics[width=0.45\textwidth]{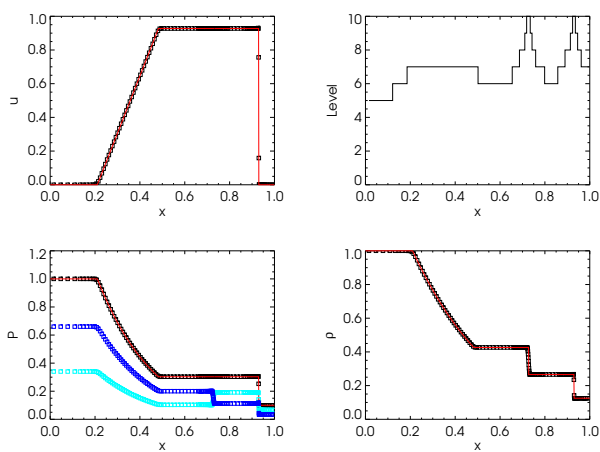}
\caption{Solution of the~\cite{sod78} shock tube problem with two temperatures. There is no diffusion and coupling of the temperatures. We plot the velocity (top left), pressure (bottom left), and density (bottom right) at $t=0.245$. The squares correspond to the result of the simulation and the red lines to the analytical solution. The light blue and dark blue squares in the pressure plot stand for the first and second species, respectively, while the black squares are for the total pressure. The level of refinement is plotted in the top right panel. }
\label{fig:sod}
\end{figure}

We set up the conditions for a~\cite{sod78} shock tube with two temperatures and without conduction or temperature coupling.
The initial left and right gas density is $\rho_{\rm L}=1$, $\rho_{\rm R}=0.125$, the first pressure $P_{\rm 1,L}=0.34$, $P_{\rm 1,R}=0.066$ (gas or ion), the second pressure $P_{\rm 2,L}=0.66$, $P_{\rm 2,R}=0.034$ (respectively CR or electron), thus $P_{\rm L}=1$ and $P_{\rm R}=0.1$, and zero velocity.
The adiabatic index for the first and second temperatures is $\gamma_{1,2}=1.4$.
The minimum level of refinement is $\ell_{\rm min}=3,$ and we refine up to level $\ell_{\rm max}=10$ wherever the relative gradient of any hydro variable is greater than $5\%$.
Figure~\ref{fig:sod} shows the result of the simulation, which exhibits extremely good agreement with the analytical solution.
At the contact discontinuity $x=0.7$, the total pressure is continuous, but not the individual pressure terms.
We reproduce the structure of a two-component shock well (see e.g.~\citealp{pfrommeretal06} for comparison).

\subsection{Shock with two temperatures and heat conduction}

\begin{figure*}
\centering \includegraphics[width=0.7\textwidth]{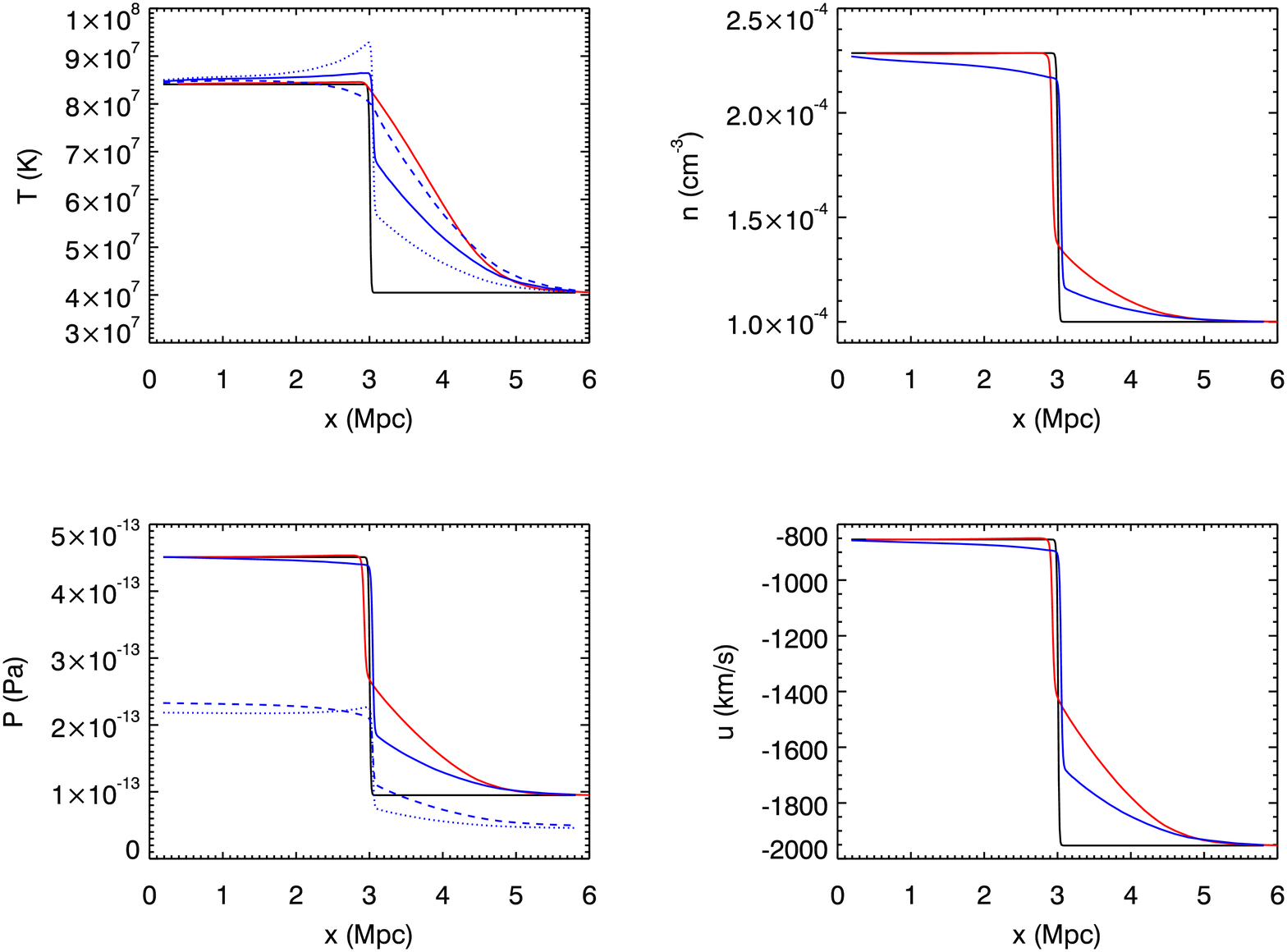}
\caption{Stationary cosmological shock experience including conduction and two-temperature coupling. The result of the two-temperature run (blue) is represented with dotted lines for the ion temperature (pressure), in dashed for the electron temperature (pressure), and solid for the average temperature (total pressure) of the gas at $t=10$ Gyr. For comparison, the results of the single temperature runs are plotted in red with conduction and in black without conduction. An electron precursor forms in the upstream part of the shock front.}
\label{fig:spitzer}
\end{figure*}

We model the formation of a 1D stationary shock typically arising at the virial radius of a cluster of galaxies.
The gas is of primordial cosmological composition with a 0.76 fraction of hydrogen and 0.24 of helium and is fully ionised.
Thus, the mean molecular weight of ions, electrons, and gas are $\mu_{\rm I}=1.22$, $\mu_{\rm E}=1.13$, and $\mu=0.59$.
The right state is with gas density $n_{\rm R}=10^{-4}$ cm$^{-3}$, gas velocity $u_{\rm R}=2000$ km s$^{-1}$ oriented in the negative $x$-direction, and ion and electron temperatures at equilibrium $T_{\rm I,R}=T_{\rm E,R}=T_{\rm R}=4\times 10^7 \, \rm K$. 
Using the Rankine-Hugoniot jump conditions, we find that the corresponding left state is $n_{\rm L}\simeq 2.3\times 10^{-4}$ cm$^{-3}$, $u_{\rm L}=875$ km s$^{-1}$, and $T_{\rm I,R}=T_{\rm E,R}=T_{\rm R}\simeq8.4\times 10^7$ K for an adiabatic index of $\gamma=5/3$.
The box length is $8$ Mpc, with a minimum and a maximum level of refinement of $\ell_{\rm min}=4$ and $\ell_{\rm max}=9,$ respectively, and with refinement triggered wherever the relative gradient of any hydro variable is larger than $10\%$.
The left and right boundaries are imposed with the initial values of left and right states.

Figure~\ref{fig:spitzer} shows the result of the simulation with the two temperatures including conduction of the electrons and the temperature coupling between ions and electrons, as well as the solution for a single temperature with or without conduction.
For the single temperature experiment without conduction, the initial left and right states are preserved at time $t=10$ Gyr.
With the addition of conduction, a thermal precursor forms $\sim 1$ Mpc ahead of the shock, which is present for both the single and the two-temperature experiments.

In the case of the two temperatures, we see that the electrons have the higher temperature value in the thermal precursor compared to the ions with a slow rise similar to the thermal precursor in the single-temperature experiment.
Ion temperature in the precursor lies below---with a sharp rise at the shock interface---because the coupling of the two temperatures is not instantaneous.
The increase in ion temperature right after the shock interface compared to the average gas temperature
is because the ions absorb the upstream kinetic energy at the shock, and electrons only lag behind and thermally recouple with the ions.
Far away from the shock front, both temperatures are at equilibrium.
The reader can refer to~\cite{zeldovich&raizer67} for the detailed description of a two-temperature shock with conduction.

\subsection{Conduction in a circular loop}

Anisotropy can be tested by the propagation of heat through a magnetic loop, here in 2D (see for comparison e.g.~\citealp{sharma&hammett07, rasera&chandran08}).
We initialised a patch of temperature of $T_{\rm }=10^8$ K wherever $2<r<3$ kpc and $165<\theta<195^\circ$ in a background gas with temperature $T_{\rm }=10^6$ K and density $n=1$ cm$^{-3}$. 
The size of the box is 10$\times$10 kpc, and we performed two runs either with a uniform grid $128^2$ or with AMR with $\ell_{\rm min}=4$ and $\ell_{\rm max}=8$ and refinement triggered if the relative gradient of pressure is larger than 20\%.
We used a Spitzer conductivity with $\kappa_\parallel=0.99\kappa_{\rm Sp}$ and $\kappa_{\rm iso}=0.01\kappa_{\rm Sp}$.

\begin{figure}
\centering \includegraphics[width=0.35\textwidth]{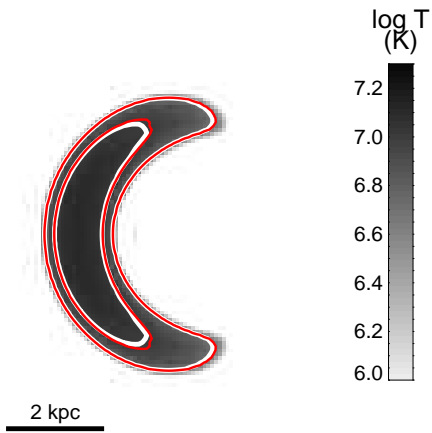}
\caption{Temperature map at $t=2$ Gyr for the heat conduction in a 2D magnetic loop without AMR (grey levels). The solution with AMR is plotted with red contours for $T=5\times 10^6$ K and $T=10^7$ K (white contours without AMR).}
\label{fig:loop2d}
\end{figure}

The result of this numerical experiment is shown in Fig.~\ref{fig:loop2d} for two runs with and without AMR.
The initial patch of temperature has diffused anisotropically along the circular-loop shape of the magnetic field lines.
The results of the AMR and of the uniform grid runs are in very good agreement, as shown by the white and red contours.

\begin{figure}
\centering \includegraphics[width=0.35\textwidth]{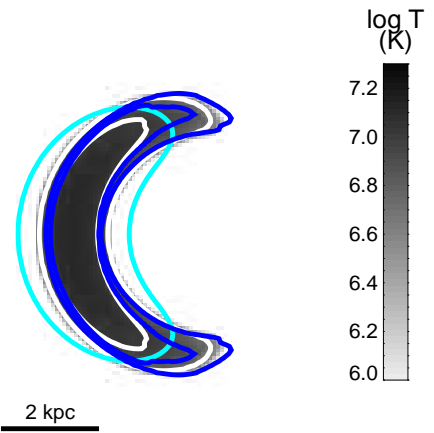}
\caption{Temperature map at $t=2 \,\rm Gyr$ for the heat conduction in a 2D magnetic loop with AMR and different fractions of perpendicular conductivity coefficients $\kappa_{\rm iso}=0.1\kappa_{\rm Sp}$ (red contours),  $\kappa_{\rm iso}=0.01\kappa_{\rm Sp}$ (grey levels and white contours), and  $\kappa_{\rm iso}=0.001\kappa_{\rm Sp}$ (blues contours). Two contours are plotted for each simulation result corresponding to temperatures $T=5\times 10^6$ K and $T=10^7$ K (except for $\kappa_{\rm iso}=0.1\kappa_{\rm Sp}$ where only $T=5\times 10^6$ K is reached). }
\label{fig:loop2dkperp}
\end{figure}

We also explored the effect of varying the perpendicular conductivity coefficient with $\kappa_{\rm iso}=[0.001,0.01,0.1]\kappa_{\rm Sp}$.
The resulting temperature maps are shown in Fig.~\ref{fig:loop2dkperp}.
As expected, the spread of temperature perpendicular to the magnetic field lines increases with the increase in the isotropic component of the diffusion coefficient.

\subsection{Sovinec test}

\begin{figure}
\centering \includegraphics[width=0.45\textwidth]{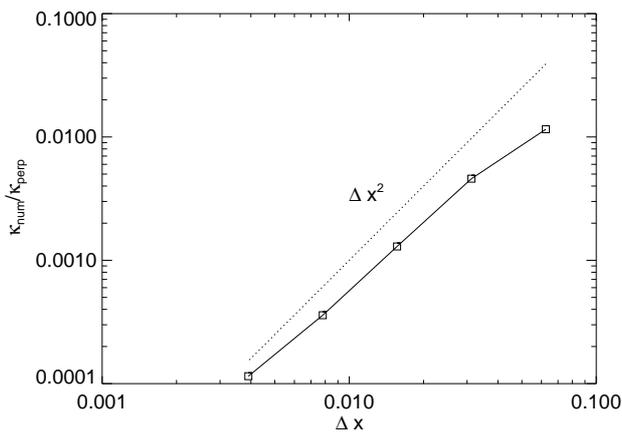}
\caption{Results of the perpendicular numerical diffusion coefficient $\kappa_{\rm num}$ as a function of resolution $\Delta x$ for the~\cite{sovinecetal04} test problem in a solid line with square symbols. The numerical diffusion scales with $\Delta x^2$ }(dotted line).
\label{fig:sovinec}
\end{figure}

\cite{sovinecetal04} designed a test to measure the numerical perpendicular diffusion $\kappa_{\rm num}$ for the anisotropic diffusion in 2D.
The energy evolution follows
\begin{equation}
\frac{\partial e} {\partial t}=-\nabla.\left(-\kappa_{\parallel}\vec b.(\vec b . \nabla) T_{\rm E} \right)-\nabla.\left(-\kappa_{\rm iso}\nabla T_{\rm E} \right) + \mathcal Q\, ,
\end{equation}
where $\mathcal Q$ is a heating source term of the form $\mathcal Q(x,y)=\mathcal Q_0\cos(\pi x) \cos (\pi y)$.
The magnetic field is chosen such that $\vec{B}.{\nabla} T_{\rm E}=0$, i.e. the parallel heat flux component is suppressed, and only the perpendicular component remains.
The analytical stationary solution at the centre is $T(0,0)=\mathcal{Q}_0/[2\pi^2(\kappa_{\rm iso}+\kappa_{\rm num})]$.
For this test $\kappa_{\rm iso}=1$ ($\kappa_{\rm Sp}=100$), $\rho=1$, $\mathcal Q_0=2\pi^2$, $B_{x}=\cos(\pi x)\sin(\pi y),$ and $B_{y}=-\sin(\pi x)\cos(\pi y)$.
The box size ranges over $[-0.5,0.5]\times[-0.5,0.5]$.
As in~\cite{rasera&chandran08}, we imposed the temperature at the boundaries of the simulation box with $T_{\rm bound}=0$.

We ran the simulation for various uniform grid resolutions from $16^2$ to $256^2$ up to the steady-state solution, and we measured the  value of $T_{\rm mea}(0,0)$. We followed~\cite{sharma&hammett07} to obtain the value of $\kappa_{\rm num}$ as follows $\kappa_{\rm num} = T_{\rm iso}(0,0)/T_{\rm mea}(0,0)-1$, where $T_{\rm iso}(0,0)$ is the  the central temperature calculated  in the isotropic limit ($\kappa_{\rm iso}=\kappa_{\rm Sp}$). This measurement is more accurate than assuming that the isotropic diffusion gives $T_{\rm iso}(0,0)=1$ exactly.
The result is shown in Fig.~\ref{fig:sovinec}, and we see that the perpendicular numerical diffusion coefficient scales quadratically with resolution $\propto \Delta x^2$ and that, even at low resolution $\Delta x=1/16$, the numerical diffusion is a factor 100 below our explicit perpendicular diffusion coefficient $\kappa_{\rm iso}$.

\subsection{Supernova explosion in a 3-phase medium: ions, electrons, and cosmic rays}

\begin{figure*}
\centering \includegraphics[width=\textwidth]{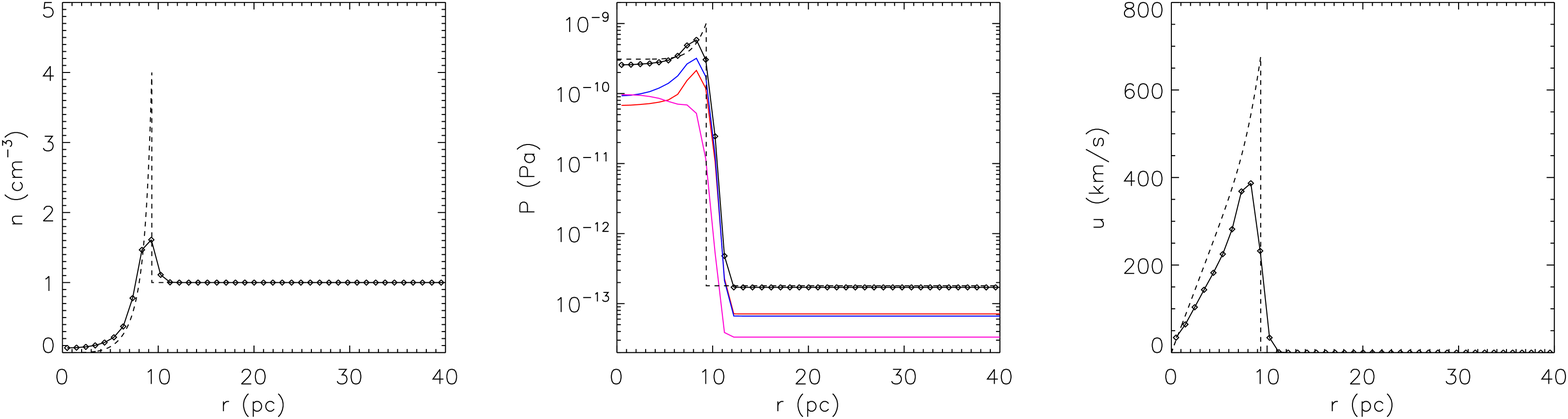}
\centering \includegraphics[width=\textwidth]{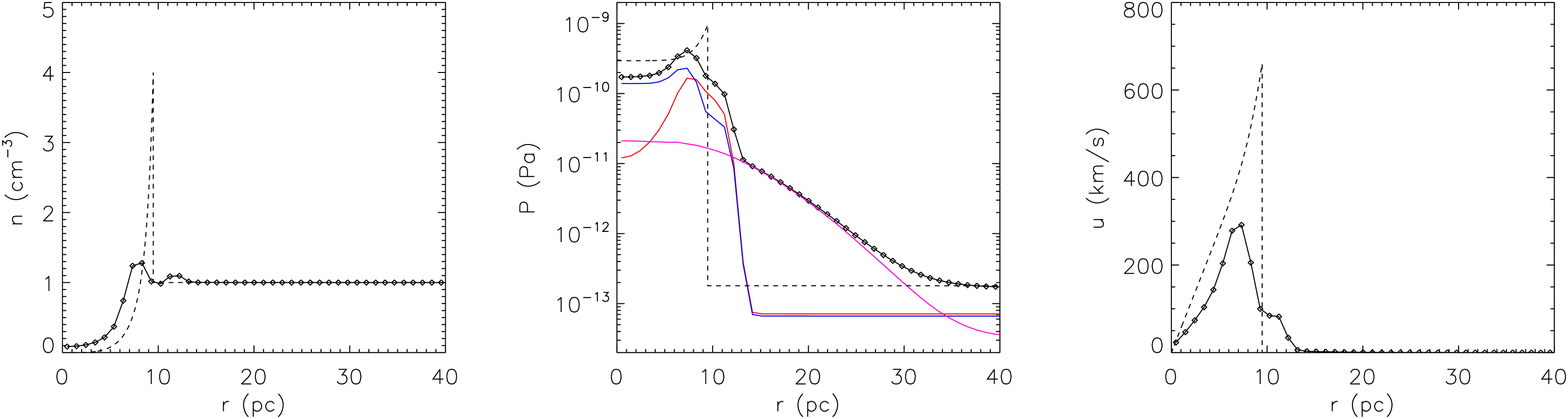}
\caption{Spherically averaged values of density (left), pressure (middle), and velocity (right) as a function of radius at $t=4$ kyr for the 3D Sedov explosion without  diffusion (top), and with isotropic diffusion (bottom). The solid lines are the result of the simulation, and the dashed line is the analytic prediction for the standard Sedov solution. The pressure is decomposed into total pressure (black), ion pressure (blue), electron pressure (red), and CR pressure (magenta). Without diffusion, the analytic solution is reproduced well. With diffusion, there are both a CR precursor and an electron precursor.}
\label{fig:sedovprof}
\end{figure*}
\begin{figure*}
\centering \includegraphics[width=\textwidth]{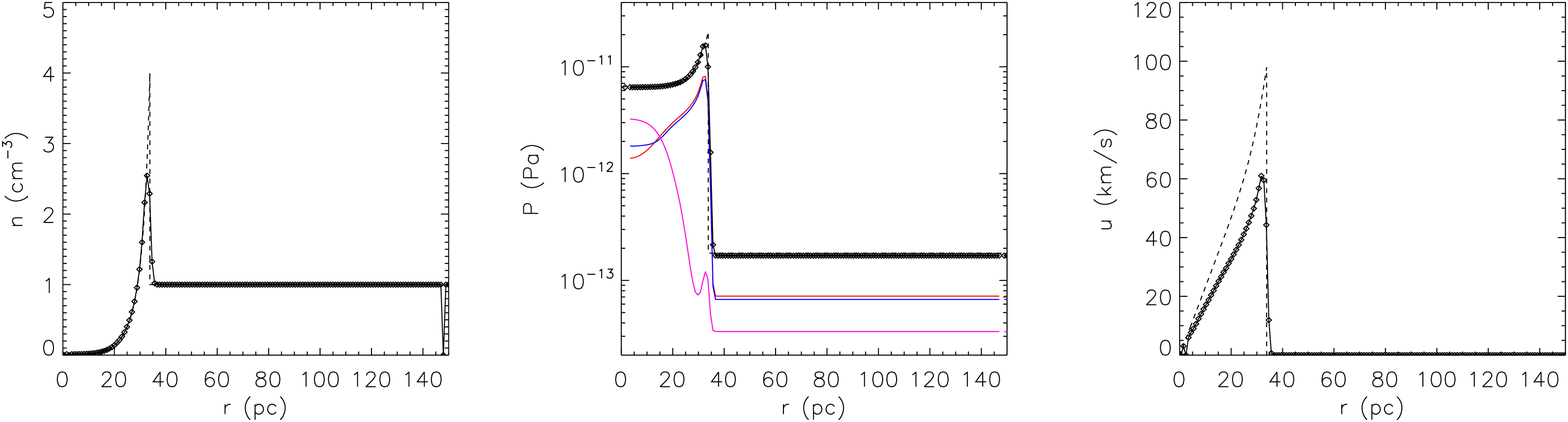}
\centering \includegraphics[width=\textwidth]{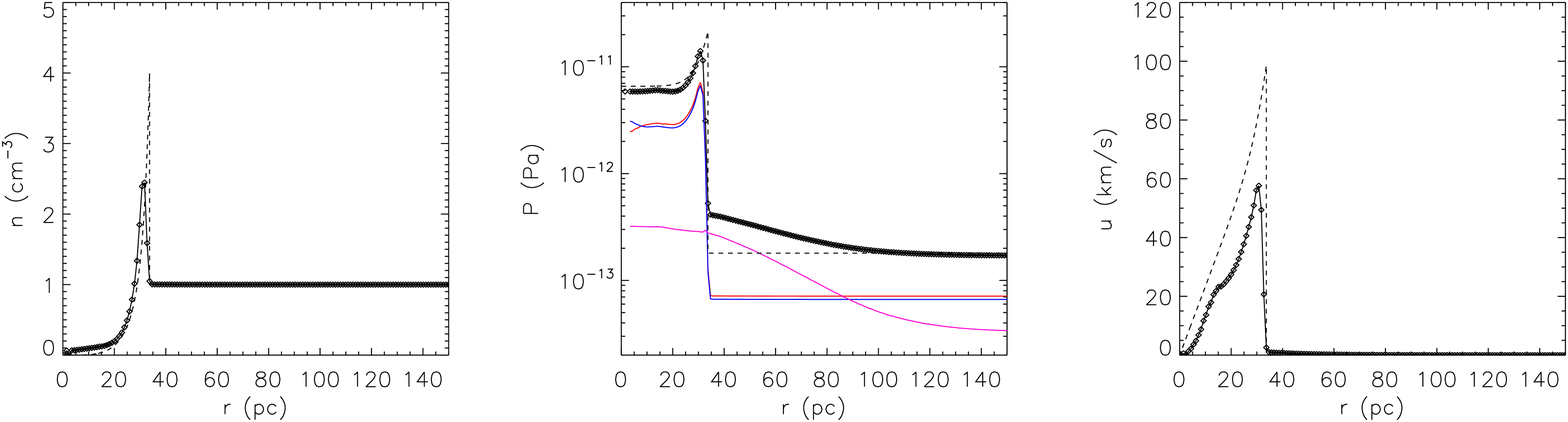}
\caption{Same as Fig.~\ref{fig:sedovprof} at $t=100$~kyr. The shock front takes over the conduction front of electrons. A CR pressure precursor is still visible in the conduction and diffusion case with a subtle effect on the Sedov shock front position, which slightly lags behind the solution without conduction and diffusion.}
\label{fig:sedovproft100}
\end{figure*}

\begin{figure*}
\centering \includegraphics[width=\textwidth]{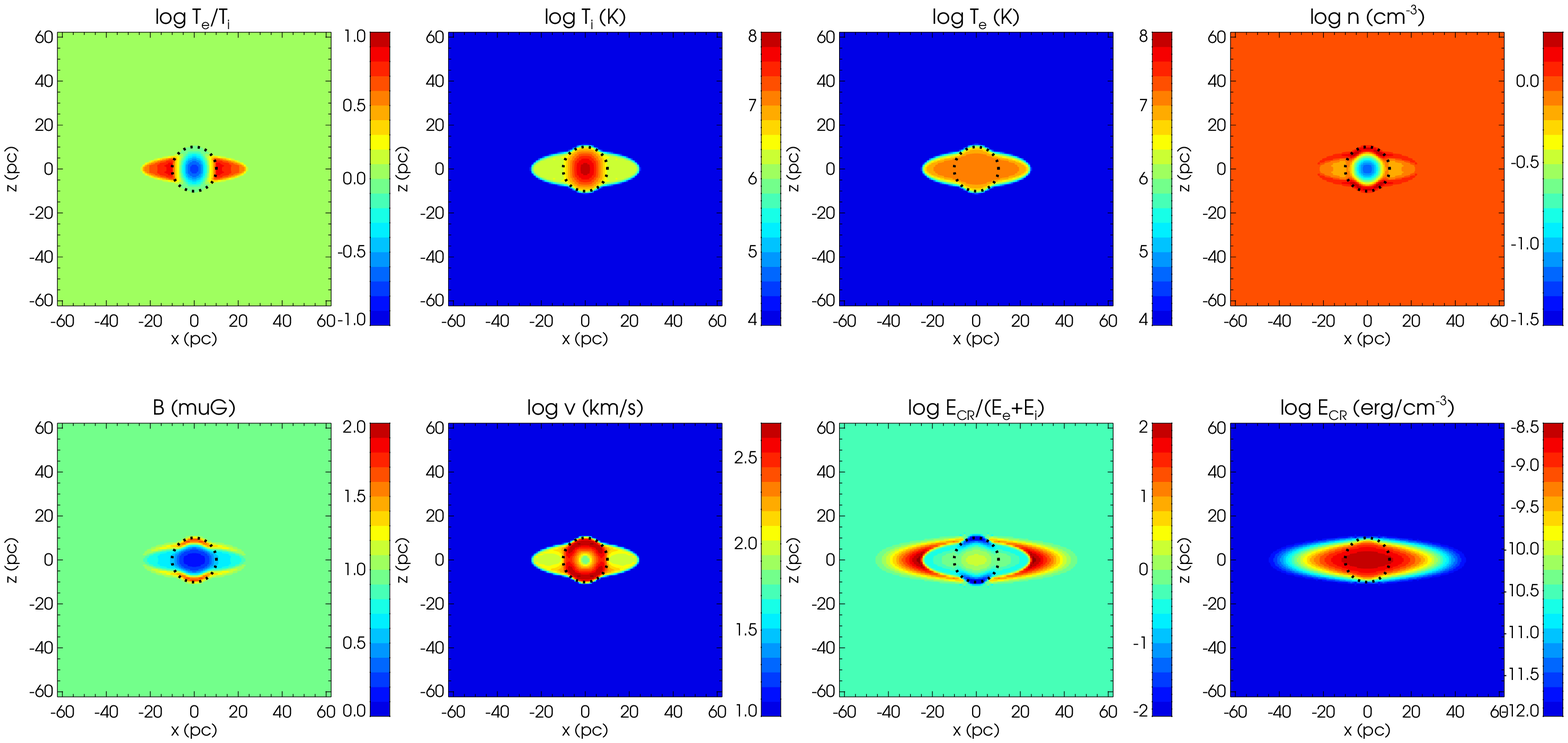}
\centering \includegraphics[width=\textwidth]{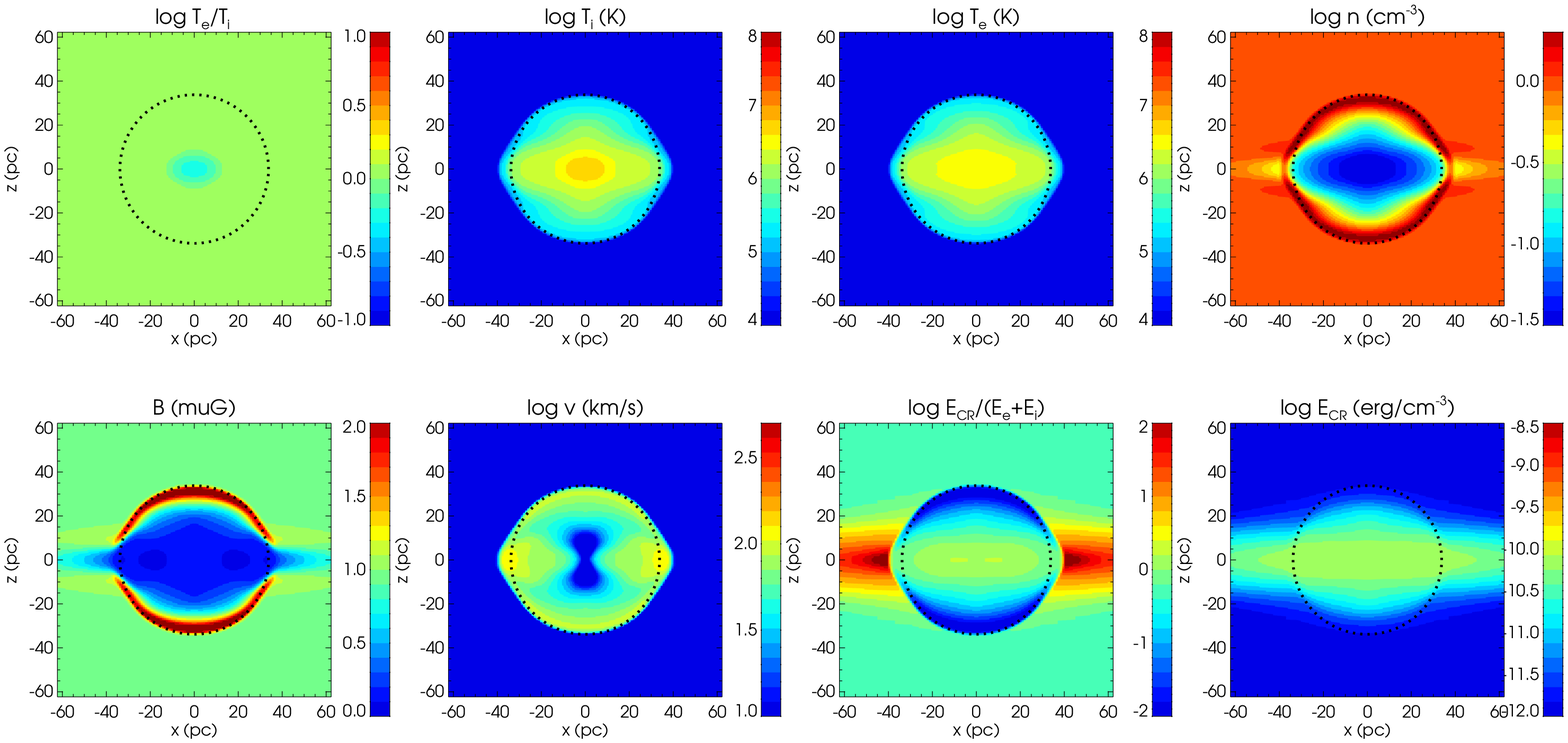}
\caption{Map cuts in the $xz$ plane for the 3D Sedov explosion including anisotropic electron heat conduction and anisotropic CR diffusion at  $t=5$ kyr (first two rows) and $t=100$ kyr (last two rows). From left to right and top to bottom: the ratio of electron to ion temperature, the ion temperature, the electron temperature, the gas density, the magnetic field amplitude, the gas velocity amplitude, the ratio of CR to internal energy, and the CR energy density. The black dotted circles are the shock front positions of the standard Sedov solution $R(t)\simeq1.1527 (E_{\rm SN}/\rho)^{1/5}t^{2/5}$. The explosion exhibits a nozzle-like heat/CR precursor in the horizontal direction due to the anisotropic nature of the diffusion (along the horizontal magnetic field).}
\label{fig:sedovani}
\end{figure*}

To fully validate the implementation of anisotropic diffusion of CRs and conduction of heat through electrons and their coupling with ions in an astrophysical context, we model the explosion of a supernova in 3D.
The explosion is triggered in a homogeneous medium of density $n=1$ cm$^{-3}$ and with temperatures of electrons and ions at equilibrium $T=10^4\, \rm K$, and the background energy density of CRs is one-third of the total energy.
We assume that the gas is totally ionised and composed of a cosmic mixture of hydrogen and helium, thus, $\mu_{\rm I}=1.22$, $\mu_{\rm E}=1.13$, and $\mu=0.59$.
The magnetic field is uniform along the $x$-axis of the box and with a magnitude of $1 \, \rm \mu G$, thus the ratio of internal to magnetic pressure is initially of $\beta=45$.
We inject in the eight central cells a total of $E_{\rm SN}=10^{51}$ erg of energy with $1/3$ in CR energy, $1/3$ in ion energy, and $1/3$ in electron energy.

The temperature of electrons is conducted at the Spitzer rate and couples with the ion temperature.
Both ion and electron adiabatic index are equal to $\gamma=5/3$. 
CRs are diffused with a diffusion coefficient $D_{\rm CR}=3\times 10^{27}$ cm$^2$ s$^{-1}$ and have an adiabatic index of $\gamma_{\rm CR} = 4/3$ corresponding to ultra-relativistic particles. 
As currently implemented, our solver could allow for several CR energy components with various diffusion coefficients and CR adiabatic indexes.
We use a box size of $500$ pc with a coarse grid of level $\ell_{\rm min}=5$ and refine up to $\ell_{\rm max}=9$ wherever the relative gradient of any hydro variable is more than $20\%$.

For the first test, we did electron and ion coupling without diffusion or conduction, and the second test was with isotropic diffusion and conduction.
Figure~\ref{fig:sedovprof} shows the result of the two simulations after time $t=4$~kyr and is compared to the analytic prediction from~\cite{sedov}.
In the absence of conduction and diffusion, the analytical solution is reproduced well, except that the shock interface is smeared by the numerical diffusion.
The electrons and ions are not yet at thermal equilibrium within the bubble.  The solution is slightly modified
with isotropic diffusion of CR energy and conduction of electron temperature.
The gas density exhibits two peaks: one ahead of the shock front position and another behind.
The ahead over-density is the product of the conduction of electrons over a few parsec, while the second over-density lags behind.
The total pressure within the bubble is decreased since the CR and electron energies have leaked out of the bubble.
Figure~\ref{fig:sedovproft100} shows the result at time $t=100$~kyr.
Again, the experiment without conduction and diffusion reproduces the analytical prediction well, and ions and electrons are at thermal equilibrium.
At this time, the Sedov shock has taken over the electron precursor, which has disappeared.
It turns out that electrons and ions are at thermal equilibrium.
However, a large amount of the CR energy has escaped the bubble, so that there is less pressure within the bubble, explaining why the shock front slightly lags behind.

The final experiment was run with anisotropic conduction of electrons and anisotropic diffusion of CRs.
Figure~\ref{fig:sedovani} shows 2D slices in the middle $xz$ plane of several quantities: temperatures, density, CR energy, velocity, and magnetic field amplitude at two different times $t=5$~kyr and $t=100$~kyr.
We see that the electron temperature and CR energy are diffused along the $x$-axis.
The ion temperature catches up with the electron temperature in the heat precursor, but it takes some time for the two species to reach thermal equilibrium.
A jet-like nozzle forms in the horizontal direction due to the anisotropic over-pressurised region, although it propagates more slowly than the genuine shock front.
Nevertheless, at later times, the explosion approaches spherical with still some degree of anisotropy in the bubble expansion.
In particular, the shock front moves faster along magnetic field lines ($x$-axis) than perpendicular to them (note the amplification of the magnetic field due to the compression along the $z$-axis).
This results in an aspherical bubble shape, which is in advance in the $x$-direction but behind in the $z$-direction (as for the $y$-direction).

\section{Conclusion}
\label{section:conclusion}

We have presented the implementation of a new solver for the anisotropic diffusion along magnetic field lines of CRs and of heat with two-temperature components, electrons, and ions in the AMR code {\sc ramses}. This solver is implicit in time and supports adaptive-time-stepping on an AMR level-by-level basis.
The implicit anisotropic diffusion scheme uses a conjugate gradient method adapted from~\cite{commerconetal11, commerconetal14}.
We tested this new implementation against several basic numerical experiments with, for some cases, typical astrophysical conditions, as well as a first simple application to a supernova explosion in a uniform medium.
This preliminary work indicates that anisotropic conduction or diffusion, together with temperature coupling, is relevant to several astrophysical problems, as already highlighted in the literature, and, in particular, may affect the propagation of remnants of supernovae.

Future applications of this new part of the {\sc ramses} code will include the study of active galactic nuclei jets in the intra-cluster medium, virial shocks in cosmological simulations of galaxy clusters, CR-driven galactic winds, supernovae explosions, and thermal instabilities in the ISM, amongst others.

\begin{acknowledgements}
We thank J. Masson for enlightening discussion of the temperature coupling.
We thank R. Teyssier, Y. Rasera, I. Parrish, T. Bogdanovich, and P. Hennebelle for useful informal discussions.
We warmly thank S. Rouberol for smoothly running the Horizon cluster on which several simulations were run.
We thank the anonymous referee for help clarifying several aspects of the paper.
This work was supported by the CNRS programmes ``Programme National de Cosmologie et Galaxies'' (PNCG) and ``Physique et Chimie du Milieu Interstellaire'' (PCMI).
BC gratefully acknowledges support from the French ANR Retour Postdoc programme
(ANR-11-PDOC-0031).
\end{acknowledgements}

\bibliographystyle{aa}
\bibliography{author}

\begin{thebibliography}{57}
\expandafter\ifx\csname natexlab\endcsname\relax\def\natexlab#1{#1}\fi

\bibitem[{{Balbus}(2000)}]{balbus00}
{Balbus}, S.~A. 2000, \apj, 534, 420

\bibitem[{{Balsara} {et~al.}(2008{\natexlab{a}}){Balsara}, {Bendinelli},
  {Tilley}, {Massari}, \& {Howk}}]{balsaraetal08sn}
{Balsara}, D.~S., {Bendinelli}, A.~J., {Tilley}, D.~A., {Massari}, A.~R., \&
  {Howk}, J.~C. 2008{\natexlab{a}}, \mnras, 386, 642

\bibitem[{{Balsara} {et~al.}(2008{\natexlab{b}}){Balsara}, {Tilley}, \&
  {Howk}}]{balsaraetal08method}
{Balsara}, D.~S., {Tilley}, D.~A., \& {Howk}, J.~C. 2008{\natexlab{b}}, \mnras,
  386, 627

\bibitem[{{Beck} \& {Krause}(2005)}]{beck&krause05}
{Beck}, R. \& {Krause}, M. 2005, Astronomische Nachrichten, 326, 414

\bibitem[{{Bogdanovi{\'c}} {et~al.}(2009){Bogdanovi{\'c}}, {Reynolds},
  {Balbus}, \& {Parrish}}]{bogdanovicetal09}
{Bogdanovi{\'c}}, T., {Reynolds}, C.~S., {Balbus}, S.~A., \& {Parrish}, I.~J.
  2009, \apj, 704, 211

\bibitem[{{Booth} {et~al.}(2013){Booth}, {Agertz}, {Kravtsov}, \&
  {Gnedin}}]{boothetal13}
{Booth}, C.~M., {Agertz}, O., {Kravtsov}, A.~V., \& {Gnedin}, N.~Y. 2013,
  \apjl, 777, L16

\bibitem[{{Brighenti} \& {Mathews}(2003)}]{brighenti&mathews03}
{Brighenti}, F. \& {Mathews}, W.~G. 2003, \apj, 587, 580

\bibitem[{{Br{\"u}ggen} {et~al.}(2005){Br{\"u}ggen}, {Ruszkowski}, \&
  {Hallman}}]{bruggenetal05}
{Br{\"u}ggen}, M., {Ruszkowski}, M., \& {Hallman}, E. 2005, \apj, 630, 740

\bibitem[{{Chi{\`e}ze} {et~al.}(1998){Chi{\`e}ze}, {Alimi}, \&
  {Teyssier}}]{chiezeetal98}
{Chi{\`e}ze}, J.-P., {Alimi}, J.-M., \& {Teyssier}, R. 1998, \apj, 495, 630

\bibitem[{{Choi} \& {Stone}(2012)}]{choi&stone12}
{Choi}, E. \& {Stone}, J.~M. 2012, \apj, 747, 86

\bibitem[{{Commer{\c c}on} {et~al.}(2014){Commer{\c c}on}, {Debout}, \&
  {Teyssier}}]{commerconetal14}
{Commer{\c c}on}, B., {Debout}, V., \& {Teyssier}, R. 2014, \aap, 563, A11

\bibitem[{{Commer{\c c}on} {et~al.}(2011){Commer{\c c}on}, {Teyssier}, {Audit},
  {Hennebelle}, \& {Chabrier}}]{commerconetal11}
{Commer{\c c}on}, B., {Teyssier}, R., {Audit}, E., {Hennebelle}, P., \&
  {Chabrier}, G. 2011, \aap, 529, A35

\bibitem[{{Courty} \& {Alimi}(2004)}]{courty&alimi04}
{Courty}, S. \& {Alimi}, J.~M. 2004, \aap, 416, 875

\bibitem[{{En{\ss}lin} {et~al.}(2011){En{\ss}lin}, {Pfrommer}, {Miniati}, \&
  {Subramanian}}]{ensslinetal11}
{En{\ss}lin}, T., {Pfrommer}, C., {Miniati}, F., \& {Subramanian}, K. 2011,
  \aap, 527, A99

\bibitem[{{Fabian}(1994)}]{fabian94}
{Fabian}, A.~C. 1994, \araa, 32, 277

\bibitem[{{Fromang} {et~al.}(2006){Fromang}, {Hennebelle}, \&
  {Teyssier}}]{fromangetal06}
{Fromang}, S., {Hennebelle}, P., \& {Teyssier}, R. 2006, \aap, 457, 371

\bibitem[{{Gaspari} \& {Churazov}(2013)}]{gaspari&churazov13}
{Gaspari}, M. \& {Churazov}, E. 2013, \aap, 559, A78

\bibitem[{{Gonz{\'a}lez} {et~al.}(2015){Gonz{\'a}lez}, {Vaytet}, {Commer{\c
  c}on}, \& {Masson}}]{gonzalezetal15}
{Gonz{\'a}lez}, M., {Vaytet}, N., {Commer{\c c}on}, B., \& {Masson}, J. 2015,
  \aap, 578, A12

\bibitem[{{G{\"u}nter} {et~al.}(2005){G{\"u}nter}, {Yu}, {Kr{\"u}ger}, \&
  {Lackner}}]{gunteretal05}
{G{\"u}nter}, S., {Yu}, Q., {Kr{\"u}ger}, J., \& {Lackner}, K. 2005, Journal of
  Computational Physics, 209, 354

\bibitem[{{Hanasz} {et~al.}(2004){Hanasz}, {Kowal}, {Otmianowska-Mazur}, \&
  {Lesch}}]{hanaszetal04}
{Hanasz}, M., {Kowal}, G., {Otmianowska-Mazur}, K., \& {Lesch}, H. 2004, \apjl,
  605, L33

\bibitem[{{Hanasz} \& {Lesch}(2003)}]{hanaszetal03}
{Hanasz}, M. \& {Lesch}, H. 2003, \aap, 412, 331

\bibitem[{{Hanasz} {et~al.}(2013){Hanasz}, {Lesch}, {Naab}, {Gawryszczak},
  {Kowalik}, \& {W{\'o}lta{\'n}ski}}]{hanaszetal13}
{Hanasz}, M., {Lesch}, H., {Naab}, T., {et~al.} 2013, \apjl, 777, L38

\bibitem[{{Jubelgas} {et~al.}(2004){Jubelgas}, {Springel}, \&
  {Dolag}}]{jubelgasetal04}
{Jubelgas}, M., {Springel}, V., \& {Dolag}, K. 2004, \mnras, 351, 423

\bibitem[{{Jubelgas} {et~al.}(2008){Jubelgas}, {Springel}, {En{\ss}lin}, \&
  {Pfrommer}}]{jubelgasetal08}
{Jubelgas}, M., {Springel}, V., {En{\ss}lin}, T., \& {Pfrommer}, C. 2008, \aap,
  481, 33

\bibitem[{{Koyama} \& {Inutsuka}(2004)}]{koyama&inutsuka04}
{Koyama}, H. \& {Inutsuka}, S.-i. 2004, \apjl, 602, L25

\bibitem[{{McNamara} \& {Nulsen}(2007)}]{mcnamara&nulsen07}
{McNamara}, B.~R. \& {Nulsen}, P.~E.~J. 2007, \araa, 45, 117

\bibitem[{{Meyer} {et~al.}(2012){Meyer}, {Balsara}, \& {Aslam}}]{meyeretal12}
{Meyer}, C.~D., {Balsara}, D.~S., \& {Aslam}, T.~D. 2012, \mnras, 422, 2102

\bibitem[{{Miniati} {et~al.}(2001){Miniati}, {Ryu}, {Kang}, \&
  {Jones}}]{miniatietal01}
{Miniati}, F., {Ryu}, D., {Kang}, H., \& {Jones}, T.~W. 2001, \apj, 559, 59

\bibitem[{{Narayan} \& {Medvedev}(2001)}]{narayan&medvedev01}
{Narayan}, R. \& {Medvedev}, M.~V. 2001, \apjl, 562, L129

\bibitem[{{Padovani} {et~al.}(2009){Padovani}, {Galli}, \&
  {Glassgold}}]{padovanietal09}
{Padovani}, M., {Galli}, D., \& {Glassgold}, A.~E. 2009, \aap, 501, 619

\bibitem[{{Parrish} \& {Quataert}(2008)}]{parrish&quataert08}
{Parrish}, I.~J. \& {Quataert}, E. 2008, \apjl, 677, L9

\bibitem[{{Parrish} {et~al.}(2009){Parrish}, {Quataert}, \&
  {Sharma}}]{parrishetal09}
{Parrish}, I.~J., {Quataert}, E., \& {Sharma}, P. 2009, \apj, 703, 96

\bibitem[{{Parrish} \& {Stone}(2005)}]{parrish&stone05}
{Parrish}, I.~J. \& {Stone}, J.~M. 2005, \apj, 633, 334

\bibitem[{{Parrish} {et~al.}(2008){Parrish}, {Stone}, \&
  {Lemaster}}]{parrishetal08}
{Parrish}, I.~J., {Stone}, J.~M., \& {Lemaster}, N. 2008, \apj, 688, 905

\bibitem[{{Pfrommer} {et~al.}(2006){Pfrommer}, {Springel}, {En{\ss}lin}, \&
  {Jubelgas}}]{pfrommeretal06}
{Pfrommer}, C., {Springel}, V., {En{\ss}lin}, T.~A., \& {Jubelgas}, M. 2006,
  \mnras, 367, 113

\bibitem[{{Piontek} \& {Ostriker}(2004)}]{piontek&ostriker04}
{Piontek}, R.~A. \& {Ostriker}, E.~C. 2004, \apj, 601, 905

\bibitem[{{Rasera} \& {Chandran}(2008)}]{rasera&chandran08}
{Rasera}, Y. \& {Chandran}, B. 2008, \apj, 685, 105

\bibitem[{{Rosner} \& {Tucker}(1989)}]{rosner&tucker89}
{Rosner}, R. \& {Tucker}, W.~H. 1989, \apj, 338, 761

\bibitem[{{Rudd} \& {Nagai}(2009)}]{rudd&nagai09}
{Rudd}, D.~H. \& {Nagai}, D. 2009, \apjl, 701, L16

\bibitem[{{Ruszkowski} \& {Begelman}(2002)}]{ruszkowski&begelman02}
{Ruszkowski}, M. \& {Begelman}, M.~C. 2002, \apj, 581, 223

\bibitem[{{Salem} \& {Bryan}(2014)}]{salem&bryan14}
{Salem}, M. \& {Bryan}, G.~L. 2014, \mnras, 437, 3312

\bibitem[{{Sedov}(1959)}]{sedov}
{Sedov}, L.~I. 1959, {Similarity and Dimensional Methods in Mechanics}

\bibitem[{{Sharma} \& {Hammett}(2007)}]{sharma&hammett07}
{Sharma}, P. \& {Hammett}, G.~W. 2007, Journal of Computational Physics, 227,
  123

\bibitem[{{Sod}(1978)}]{sod78}
{Sod}, G.~A. 1978, Journal of Computational Physics, 27, 1

\bibitem[{{Sovinec} {et~al.}(2004){Sovinec}, {Glasser}, {Gianakon}, {Barnes},
  {Nebel}, {Kruger}, {Schnack}, {Plimpton}, {Tarditi}, {Chu}, \& {Nimrod
  Team}}]{sovinecetal04}
{Sovinec}, C.~R., {Glasser}, A.~H., {Gianakon}, T.~A., {et~al.} 2004, Journal
  of Computational Physics, 195, 355

\bibitem[{{Spitzer}(1956)}]{spitzer56}
{Spitzer}, L. 1956, {Physics of Fully Ionized Gases}

\bibitem[{{Spitzer} \& {Tomasko}(1968)}]{spitzer&tomasko68}
{Spitzer}, Jr., L. \& {Tomasko}, M.~G. 1968, \apj, 152, 971

\bibitem[{{Strong} {et~al.}(2007){Strong}, {Moskalenko}, \&
  {Ptuskin}}]{strongetal07}
{Strong}, A.~W., {Moskalenko}, I.~V., \& {Ptuskin}, V.~S. 2007, Annual Review
  of Nuclear and Particle Science, 57, 285

\bibitem[{{Teyssier}(2002)}]{teyssier02}
{Teyssier}, R. 2002, \aap, 385, 337

\bibitem[{{Teyssier} {et~al.}(1998){Teyssier}, {Chi{\`e}ze}, \&
  {Alimi}}]{teyssieretal98}
{Teyssier}, R., {Chi{\`e}ze}, J.-P., \& {Alimi}, J.-M. 1998, \apj, 509, 62

\bibitem[{{Teyssier} {et~al.}(2006){Teyssier}, {Fromang}, \&
  {Dormy}}]{teyssieretal06}
{Teyssier}, R., {Fromang}, S., \& {Dormy}, E. 2006, Journal of Computational
  Physics, 218, 44

\bibitem[{{Tilley} {et~al.}(2006){Tilley}, {Balsara}, \& {Howk}}]{tilleyetal06}
{Tilley}, D.~A., {Balsara}, D.~S., \& {Howk}, J.~C. 2006, \mnras, 371, 1106

\bibitem[{{Uhlig} {et~al.}(2012){Uhlig}, {Pfrommer}, {Sharma}, {Nath},
  {En{\ss}lin}, \& {Springel}}]{uhligetal12}
{Uhlig}, M., {Pfrommer}, C., {Sharma}, M., {et~al.} 2012, \mnras, 423, 2374

\bibitem[{{van Leer}(1979)}]{vanleer79}
{van Leer}, B. 1979, Journal of Computational Physics, 32, 101

\bibitem[{{Wong} \& {Sarazin}(2009)}]{wong&sarazin09}
{Wong}, K.-W. \& {Sarazin}, C.~L. 2009, \apj, 707, 1141

\bibitem[{{Yokoyama} \& {Shibata}(2001)}]{yokoyama&shibata01}
{Yokoyama}, T. \& {Shibata}, K. 2001, \apj, 549, 1160

\bibitem[{{Zel'dovich} \& {Raizer}(1967)}]{zeldovich&raizer67}
{Zel'dovich}, Y.~B. \& {Raizer}, Y.~P. 1967, {Physics of shock waves and
  high-temperature hydrodynamic phenomena}

\end{thebibliography}

\end{document}